\documentstyle[12pt]{article}
\newcommand{\bea}{\begin{eqnarray}}
\newcommand{\eea}{\end{eqnarray}}
\newcommand{\be}{\begin{equation}}
\newcommand{\ee}{\end{equation}}

\newcommand{\nn}{\nonumber}

\begin{document}


\begin{center}
{\bf MESONS OF THE $f_0$-FAMILY IN THE PROCESSES
$\pi\pi\to\pi\pi,K\overline{K},\eta\eta$\\ UP TO 1.9 GEV}
\footnote{
This work was supported by the Grant Program of Plenipotentiary of the
Slovak Republic at JINR.  Yu.S. and M.N. were supported in part by the
Slovak Scientific Grant Agency, Grant VEGA No. 2/7175/20; and D.K., by
Grant VEGA No. 2/5085/99.}

\bigskip

\underline{Yu.S.~Surovtsev}$^a$, D.~Krupa$^b$, M.~Nagy$^b$\\
\end{center}
$~~~~~~~^a$ {\it Bogoliubov Laboratory of Theoretical Physics, JINR, Dubna
141 980, Russia}\\ $~~~~~~~~^b$ {\it Institute of Physics, SAS,
D\'ubravsk\'a cesta 9, 842 28 Bratislava, Slovakia} \begin{abstract}
\baselineskip 10pt In combined 2- and 3-channel analyses of experimental
data on the coupled processes $\pi\pi\to\pi\pi,K\overline{K},\eta\eta$ in
the channel with $I^GJ^{PC}=0^+0^{++}$, various scenarios of these
reactions(with different numbers of resonances) are considered. In a
model-independent approach, confirmation of the $\sigma$-meson below 1 GeV
and definite indications of the QCD nature of other $f_0$ resonances are
obtained. The conclusion on the linear realization of chiral symmetry is
drawn. \end{abstract}

\noindent{\bf 1.}
In recent years, the problem of scalar mesons is the most discussed
topic in the light hadron spectroscopy. This is related to an important
role played by lightest scalars in the hadronic dynamics. For example, a
recent discovery of the $\sigma$-meson below 1 GeV \cite{PDG-02}-
\cite{Li-Zou-Li} leads to an important conclusion about the linear
realization of chiral symmetry.
The $f_0$ mesons are most direct carrier of information about the QCD
vacuum. Therefore, every step in understanding the nature of the $f_0$
mesons is especially important.

Obviously, it is important to have a model-independent
information on investigated states and on their QCD nature. It can be
obtained only on the basis of the first principles (analyticity and
unitarity) immediately applied to analyzing experimental data. Earlier, we
have proposed the method for 2- and 3-channel resonances \cite{KMS-nc96}.
We outline this further.

\noindent{\bf 2. Two- and Three-Coupled-Channel Formalism.}
We consider the processes $\pi\pi\to\pi\pi,K\overline{K}$ in the
2-channel approach and, adding the process $\pi\pi\to\eta\eta$, we use the
3-channel one. Therefore, the $S$-matrix is determined on the 4- and
8-sheeted Riemann surfaces, respectively. The matrix elements
$S_{\alpha\beta}$, where $\alpha,\beta=1(\pi\pi), 2(K\overline{K}),
3(\eta\eta)$, have the right-hand cuts along the real axis of the $s$
complex plane ($s$ is the invariant total energy squared), starting with
$4m_\pi^2$, $4m_K^2$, and $4m_\eta^2$, and the left-hand cuts. The
Riemann-surface sheets are numbered according to the signs of analytic
continuations of the channel momenta $$k_1=(s/4-m_\pi^2)^{1/2},~~~~~~
k_2=(s/4-m_K^2)^{1/2},~~~~~~k_1=(s/4-m_\pi^2)^{1/2}$$ as follows: in
the 2-channel case, signs $({\mbox{Im}}k_1,{\mbox{Im}}k_2)=++,-+,--,+-$
correspond to sheets I, II, III, IV; in the 3-channel case, signs
$({\mbox{Im}}k_1,{\mbox{Im}}k_2,{\mbox{Im}}k_3)=+++,-++,--+,+-+,+--,---,-+
-, ++-$ correspond to sheets I, II,$\cdots$, VIII.

The resonance representations on the Riemann surface are obtained with the
help of the formulae (see, ref. \cite{KMS-nc96}, table 1), expressing
analytic continuations of the matrix elements to unphysical sheets in
terms of those on sheet I -- $S_{\alpha\beta}^I$ that have only zeros
(beyond the real axis) corresponding to resonances. These formulae
immediately give the resonance representation by poles and zeros on the
Riemann surfaces if one starts from resonance zeros on sheet I. In the
2-channel approach, we obtain 3 types of resonances described by a pair of
conjugate zeros on sheet I: ({\bf a}) in $S_{11}$, ({\bf b}) in $S_{22}$,
({\bf c}) in each of $S_{11}$ and $S_{22}$.

In the 3-channel case, we obtain 7 types of resonances corresponding to
conjugate resonance zeros on sheet I of ({\bf a}) $S_{11}$; ({\bf b})
$S_{22}$; ({\bf c}) $S_{33}$; ({\bf d}) $S_{11}$ and $S_{22}$; ({\bf e})
$S_{22}$ and $S_{33}$; ({\bf f}) $S_{11}$ and $S_{33}$; and ({\bf g})
$S_{11}$, $S_{22}$, and $S_{33}$. For example, the arrangement of poles
corresponding to a ({\bf g}) resonance is: each sheet II, IV, and VIII
contains a pair of conjugate poles at the points that are zeros on sheet
I; each sheet III, V, and VII contains two pairs of conjugate poles; and
sheet VI contains three pairs of poles.

The resonance of each type is represented by a pair of complex-conjugate
clusters (of poles and zeros on the Riemann surface) of size typical of
strong interactions. The cluster kind is related to the state nature. For
example, a 2-channel resonance, coupled relatively more strongly to the
$\pi\pi$ channel than to the $K\overline{K}$ one, is described by the
({\bf a}) cluster; in the opposite case, by the ({\bf b}) one (the state
with the dominant $s{\bar s}$ component); the flavour singlet ({\it e.g.},
glueball) must be represented by the ({\bf c}) cluster.

Furthermore, according to the type of pole clusters, we can distinguish, in a
model-independent way, a bound state of colourless particles ({\it e.g.},
$K\overline{K}$ molecule) and a $q{\bar q}$ bound state \cite{KMS-nc96,MP-93}.
Just as in the 1-channel case, the existence of a particle bound-state means
the presence of a pole on the real axis under the threshold on the physical
sheet, so in the 2-channel case, the existence of a particle bound-state in
channel 2 ($K\overline{K}$ molecule) that, however, can decay into channel
1 ($\pi\pi$ decay), would imply the presence of a pair of complex conjugate
poles on sheet II under the second-channel threshold without an accompaniment
of the corresponding shifted pair of poles on sheet III. Namely, according
to this test, earlier, the interpretation of the $f_0(980)$ state as a
$K\overline{K}$ molecule has been rejected. In the 3-channel case,
the bound-state in channel 3 ($(\eta\eta)$) that, however, can decay into
channels 1 ($\pi\pi$ decay) and 2 ($K\overline{K}$ decay), is represented by
the pair of complex conjugate poles on sheet II and by a shifted poles on
sheet III under the $\eta\eta$ threshold without an accompaniment of the
corresponding poles on sheets VI and VII.

Further we use the uniformization procedure, {\it i.e.}, in the uniformizing
variable we take into account the branch points corresponding to the
thresholds of the coupled channels, and the one related to the crossing
channels. On the uniformization plane, the pole-cluster representation of
a resonance is a good one.

For a combined analysis of data on coupled processes it is
convenient to use the Le Couteur-Newton relations \cite{LN} expressing the
$S$-matrix elements of all coupled processes in terms of the Jost matrix
determinant $d(k_1,k_2,\cdots)$ being the real analytic function with the
only square-root branch-points at $k_i=0$.

\noindent{\bf 2.1. Two-Channel Approach.} Here we take into account the
right-hand branch-points at $4m_\pi^2$ and $4m_K^2$ and also the left-hand
one at $s=0$ in $S_{11}$, using the uniformizing variable
\be
v=\frac{m_K\sqrt{s-4m_\pi^2}+m_\pi\sqrt{s-4m_K^2}}{\sqrt{s(m_K^2-m_\pi^2)}},
\ee
which maps the 4-sheeted Riemann surface onto the $v$-plane, divided into
two parts by a unit circle centered at the origin. Sheets I (II), III (IV)
are mapped onto the exterior (interior) of the unit disk on the upper and
lower $v$-half-plane, respectively. The physical region extends from the
point $i$ on the imaginary axis ($\pi\pi$ threshold) along the unit circle
clockwise in the 1st quadrant to the point 1 on the real axis
($K\overline{K}$ threshold) and then along the real axis to the point
$b=\sqrt{(m_K+m_\pi)/(m_K-m_\pi)}$ into which $s=\infty$ is mapped on the
$v$-plane. The intervals $(-\infty,-b],[-b^{-1},b^{-1}],[b,\infty)$ on the
real axis are the images of the corresponding edges of the left-hand cut
of the $\pi\pi$-scattering amplitude. The ({\bf a}) resonance is
represented in $S_{11}$ by two pairs of poles on the images of sheets II
and III, symmetric to each other with respect to the imaginary axis, and
by zeros, symmetric to these poles with respect to the unit circle. On the
$v$-plane, the Le Couteur-Newton relations are
\be
S_{11}=\frac{d(-v^{-1})}{d(v)},\qquad S_{22}=\frac{d(v^{-1})}{d(v)},\qquad
S_{11}S_{22}-S_{12}^2=\frac{d(-v)}{d(v)}.
\ee
The $d(v)$-function is taken as
$~d=d_B d_{res}$ where the resonance contribution $d_{res}(v)$ is
\be
d_{res} =v^{-M}\prod_{n=1}^{M} (1-v_n^* v)(1+v_n v)
\ee
($M$ is the number
of pairs of conjugate zeros); $d_B$ is the $K\overline{K}$ background. On
the $v$-plane, $S_{11}$ has no cuts, however, $S_{12}$ does have the cut
arising from the left-hand cut on the $s$-plane. This left-hand cut is
neglected in the Riemann-surface structure, and the contribution on this
cut is taken into account in the $K\overline{K}$ background \cite{PRD-01}:
$$d_B=v^{-4}(1-pv)^4(1+p^*v)^4$$ where $p$ is the position of zero on the
unit circle.

We analyzed in combined way the data on the processes
$\pi\pi\to\pi\pi,K\overline{K}$ in the channel with $I^GJ^{PC}=0^+0^{++}$.
For the $\pi\pi$-scattering, the data from the threshold to 1.89 GeV are
taken from ref. \cite{Hyams}; below 1 GeV, from many works \cite{Hyams}-
\cite{Zylber}.
For $\pi\pi\to K\overline{K}$, practically all the accessible data are
used \cite{Wetzel}. We consider 4 variants with the following states: \\
\underline{Variant 1}: The $f_0(600)$ and $f_0 (980)$) with the type ({\bf
a}) clusters, and $f_0(1500)$, of the type ({\bf c}); \\
\underline{Variant 2}: The same three resonances $+$ the ${f_0}(1370)$ of
the type ({\bf b});\\ \underline{Variant 3}: The $f_0(600)$, $f_0 (980)$)
and $f_0 (1500)$ $+$ the ${f_0}(1710)$ of the type ({\bf b});\\
\underline{Variant 4}: All the five resonances of the indicated types.

The other possibilities of the representation of these states are rejected
by our analysis. The $\pi\pi$-scattering data are described satisfactorily
from the threshold to 1.89 GeV in all four variants. For $\pi\pi\to
K\overline{K}$, the description ranges are slightly different for various
variants and extend from the threshold to $\sim$ 1.4 GeV for variant 1, to
$\sim$ 1.46 GeV for variant 2, and to $\sim$ 1.5 GeV for variants 3 and 4.
We have obtained the following quality of fits to the data (the total
$\chi^2/\mbox{ndf}$ for both processes, the number of adjusted parameters,
p): Variant 1 -- 1.98, 17, 0.954381+0.29859i; Variant 2 -- 2.45, 21,
0.97925+0.202657i; Variant 3 -- 1.76, 21, 0.954572+0.29798i; Variant 4 --
2.59, 25, 0.982091+0.188405i.

Two best variants 1 and 3 are both without the ${f_0}(1370)$.
In table\ref{tab:clusters4} we show the obtained pole clusters of
resonances on the complex energy plane ($\sqrt{s_r}={\rm E}_r-i\Gamma_r$)
for variant 4 with all five states that are in the PDG tables
\cite{PDG-02}. \begin{table}[ht] \centering \caption{Pole clusters for
resonances in variant 4.} \vskip0.3truecm \begin{tabular}{|c|c|c|c|c|}
\hline \multicolumn{2}{|c|}{Sheet} & II & III & IV \\ \hline {$f_0 (600)$}
& {E, MeV} & 600$\pm$16 & 715$\pm$17 & {}\\ {} & {$\Gamma$, MeV} &
605$\pm$28 & 59$\pm$6 & {} \\ \hline {$f_0 (980)$} & {E, MeV} & 985$\pm$5~
& 984$\pm$18 & {}\\ {} & {$\Gamma$, MeV} & 27$\pm$8 & 210$\pm$22 & {} \\
\hline {$f_0 (1370)$} & {E, MeV} & {} & 1310$\pm$22~ & 1320$\pm$20~ \\ {}
& {$\Gamma$, MeV} & {} & 410$\pm$29 & 275$\pm$25 \\ \hline {$f_0 (1500)$}
& {E, MeV} & 1528$\pm$22~ & 1490$\pm$30~~1510$\pm$20 & 1510$\pm$21~ \\ {}
& {$\Gamma$, MeV} & 385$\pm$25 & ~220$\pm$24~~~~370$\pm$30 & 308$\pm$30 \\
\hline {$f_0 (1710)$} & {E, MeV} & {} & 1700$\pm$25~~ & 1700$\pm$20~ \\ {}
& {$\Gamma$, MeV} & {} & 86$\pm$16 & 115$\pm$20 \\ \hline \hline
\end{tabular} \label{tab:clusters4} \end{table}

The coupling constants of states with the $\pi\pi$ ($g_1$) and
$K\overline{K}$ ($g_2$) systems are calculated through the residues of the
amplitudes at the pole on sheet II -- for the ({\bf a}) and ({\bf c})
resonances, and on sheet IV -- for the ({\bf b}) resonances. Taking
$$S_{ii}=1+2i\rho_iT_{ii},~~~~~~S_{12}=2i\sqrt{\rho_1\rho_2} T_{12}$$ with
$\rho_i=\sqrt{(s-4m_i^2)/s}$ and the resonance part of the amplitude as
$$T_{ij}^{res}=\sum_r g_{ir}g_{rj}D_r^{-1}(s)$$ with $D_r(s)$ being an
inverse propagator ($D_r(s)\propto s-s_r$), we obtain (in GeV units): for
$f_0(600)$: $g_1=0.652\pm 0.065$ and $g_2=0.724\pm 0.1$, for $f_0(980)$:
$g_1=0.167\pm 0.05$ and $g_2=0.445\pm 0.031$, for $f_0(1370)$:
$g_1=0.116\pm 0.03$ and $g_2=0.99\pm 0.05$, for $f_0(1500)$: $g_1=0.657
\pm 0.113$ and $g_2=0.666\pm 0.15$.

We can see that $f_0(980)$ and ${f_0}(1370)$ are coupled
essentially more strongly to the $K\overline{K}$ system than to the
$\pi\pi$ one, which implies the dominant $s{\bar s}$ component in
these states.
The $f_0(1500)$ has approximately equal coupling constants with the
$\pi\pi$ and $K\overline{K}$ systems, which could point to
its dominant glueball component. The $f_0(1710)$ is represented by the
cluster corresponding to a dominant $s{\bar s}$ component.

Let us also present the calculated scattering lengths, first, for the
$K\overline{K}$ scattering: \\ \hspace*{3.cm}$a_0^0 = -1.25\pm
0.11+(0.65\pm 0.09)i,~ [m_{\pi^+}^{-1}]; ~~~~~~~~~~~~~({\rm variant 1}),$
\\ \hspace*{3.cm}$a_0^0= -1.548\pm 0.13+(0.634\pm 0.1)i,~
[m_{\pi^+}^{-1}]; ~~~~~~~~~~~~({\rm variant 2}),$ \\ \hspace*{3.cm}$a_0^0
= -1.19\pm 0.08+(0.622\pm 0.07)i,~ [m_{\pi^+}^{-1}]; ~~~~~~~~~~~~({\rm
variant 3}),$\\ \hspace*{3.cm}$a_0^0 = -1.58\pm 0.12+(0.59\pm 0.1)i,~
[m_{\pi^+}^{-1}]; ~~~~~~~~~~~~~~~({\rm variant 4}).$ \\ Variants 2 and 4
include $f_0(1370)$. We can see that ${\rm Re}~a_0^0(K\overline{K})$ is
very sensitive to whether this state exists or not.

In Table~\ref{tab:pipi.length}, we compare our results for the $\pi\pi$
scattering length with the results of some other theoretical and
experimental works. \begin{table}[htb] \centering \caption{}
\vskip0.3truecm \begin{tabular}{|c|l|l|} \hline $a_0^0, ~m_{\pi^+}^{-1}$ &
~~~~~~~References & ~~~~~~~~~~~~~~~~~Remarks \\ \hline $0.27\pm 0.06$ (1)&
our paper & model-independent approach \\ $0.267\pm 0.07$ (2)&{}&{}\\
$0.28\pm 0.05$ (3)&{}&{}\\ $0.27\pm 0.08$ (4)&{}&{}\\ \hline $0.26\pm
0.05$ & L. Rosselet et al.\cite{Zylber} & analysis of the decay $K\to\pi\pi
e\nu$ \\ {} & {} & using Roy's model\\ \hline $0.24\pm 0.09$ & A.A.
Bel'kov et al.\cite{Zylber} & analysis of $\pi^- p\to\pi^+\pi^-n$ \\ {} &
{} & using the effective range formula\\
 \hline $0.16$ & S.
Weinberg \cite{Weinberg} &
current algebra (non-linear $\sigma$-model) \\
\hline $0.20$ & J. Gasser,
H. Leutwyler \cite{Gasser} & one-loop
corrections, non-linear\\ {} & {}
& realization of chiral symmetry \\ \hline
$0.217$ & J. Bijnens et
al.\cite{Bijnens} & two-loop corrections, non-
linear\\ {} & {} &
realization of chiral symmetry  \\ \hline $0.26$
& M.K. Volkov
\cite{Volkov} & linear realization of chiral
symmetry \\ \hline $0.28$ &
A.N.Ivanov, N.Troitskaya \cite{Ivanov-Tr} & a
variant of chiral theory
with\\ {} & {} & linear realization of chiral
symmetry \\ \hline
\end{tabular}
\label{tab:pipi.length}
\end{table}

Our results correspond to the linear realization of chiral symmetry.

\noindent{\bf 2.2. Three-Channel Approach.} We will take into account also
the branch-point of the $\eta\eta$ threshold; {\it i.e.}, in this case, we
have the 8-sheeted Riemann surface. However, it is impossible to map that
surface onto a plane with the help of a simple function.
With the help of a simple mapping, a function, determined on the 8-sheeted
Riemann surface, can be uniformized only on torus. This is
unsatisfactory for our purpose. Therefore, we neglect the influence of the
$\pi\pi$-threshold branch point (however, unitarity on the $\pi\pi$-cut is
taken into account). An approximation like that means the consideration of
the nearest to the physical region semi-sheets of the Riemann surface. In
fact, we construct a 4-sheeted model of the initial Riemann surface
approximating it in accordance with our approach of a consistent account of
the nearest singularities on all the relevant sheets. The uniformizing
variable can be chosen as
\be
w=\frac{k_2+k_3}{\sqrt{m_\eta^2-m_K^2}}.
\ee
It maps
our model of the 8-sheeted Riemann surface onto the $w$-plane divided into
two parts by a unit circle centered at the origin. The sheets I (III), II
(IV), V (VII) and VI (VIII) are mapped onto the exterior (interior) of the
unit disk in the 1st, 2nd, 3rd and 4th quadrants, respectively. The
physical region extends from the point $w_\pi$ on the imaginary axis
($\pi\pi$ threshold, $|w_\pi|>1$) down this axis to the point i on the
unit circle ($K\overline{K}$ threshold), further along the unit circle
clockwise in the 1st quadrant to point 1 on the real axis ($\eta\eta$
threshold) and then along the real axis to $\infty$. The type ({\bf a})
resonance is represented in $S_{11}$ by the pole on the image of the sheet
II, of the sheet III, VI and VII and by zeros, symmetric to these poles
with respect to the imaginary axis. Here the left-hand cuts are neglected
in the Riemann-surface structure, and contributions on these cuts will be
taken into account in the background.

On the $w$-plane, the Le Couteur-Newton relations are
\bea
S_{11}=\frac{d^* (-w^*)}{d(w)},~~~~~S_{22}=\frac{d(-w^{-1})}{d(w)},~~~~~
S_{33}=\frac{d(w^{-1})}{d(w)},\nn\\
~~~~~~~~~~~~~~S_{11}S_{22}-S_{12}^2=\frac{d^*({w^*}^{-1})}{d(w)},~~~~~~~~~
S_{11}S_{33}-S_{13}^2=\frac{d^* (-{w^*}^{-1})}{d(w)}.
\eea
Taking the $d$-function
as $d=d_B d_{res}$ where $d_B$, describing the background, is
\be
d_B=\mbox{exp}[-i\sum_{n=1}^{3}k_n(\alpha_n+i\beta_n)].
\ee
The resonance part is
\be
d_{res}(w)=w^{-\frac{M}{2}}\prod_{r=1}^{M}(w+w_{r}^*)
\ee
where $M$ is the number of resonance zeros.

We analyzed the data on three processes
$\pi\pi\to\pi\pi,K\overline{K},\eta\eta$. The $|S_{13}|^2$ data for
$\pi\pi\to\eta\eta$ from the threshold to 1.72 GeV are taken from ref.
\cite{Binon}. Here we consider a variant with all five resonances.
Furthermore, in view of a big amount of fitting parameters in this case,
we are forced to reduce their amount assuming that single poles on the
Riemann sheets are the result of the simple multichannel Breit--Wigner
form. Then for the
$\pi\pi$-scattering, we obtain a satisfactory description from $\sim$ 0.4
GeV to 1.89 GeV ($\chi^2/\mbox{ndf}\approx 1.5$). Taking into account the
the $\eta\eta$-threshold branch-point, we have considerably improved the
description of the phase shift for the process $\pi\pi\to K\overline{K}$
($\chi^2/\mbox{ndf}\approx 2.2$) by comparison with the 2-cannel case. The
total $\chi^2/\mbox{ndf}$ for all three processes is 1.74; the number of
adjusted parameters is 40. The background parameters (in GeV$^{-1}$ units)
are $\alpha_1=1.37, \beta_1=0, \alpha_2=-1.43,\beta_2=0.2,\beta_3=0.928$.
In Table~\ref{tab:cluster34}, the obtained pole clusters for resonances
are shown (poles on sheets V and VI, corresponding to the $f_0 (1370)$ and
$f_0 (1710)$, are of the 2nd order; the $f_0 (1500)$ cluster has a pole of
the 3rd order on sheet VI and a pole of the 2nd order on each of sheets
III, V and VII). \begin{table}[ht] \centering \caption{Pole clusters for
considered resonances on the 8-sheeted Riemann surface.} \vskip0.3truecm
\begin{tabular}{|c|c|c|c|c|c|c|c|c|} \hline \multicolumn{2}{|c|}{Sheet} &
II & III & IV & V & VI & VII & VIII \\ \hline {$f_0 (600)$} & {E, MeV} &
683 & 672 & {} & {} & 627 & 638 & {} \\ {} & {$\Gamma$, MeV} & 600 & 600 &
{} &{} & 600 & 600 & {}\\ \hline {$f_0(980)$} & {E, MeV} & 1004 & 963 & {}
& {} & {} & {} & {}\\ {} & {$\Gamma$, MeV} & 39 & 76 & {} & {} & {} & {} &
{} \\ \hline {$f_0 (1370)$} & {E, MeV} & {} & 1380 & 1380 & 1367 & 1361 &
1380 & 1380 \\ {} & {$\Gamma$, MeV} & {} & 116 & 134 & 260 & 253 & 134 &
116 \\ \hline {$f_0 (1500)$} & {E, MeV} & 1505 & 1505 & 1505 & 1501 & 1515
& 1491 & 1505 \\ {} & {$\Gamma$, MeV} & 326 & 260 & 250 & 127 & 172 & 116
& 110 \\ \hline {$f_0 (1710)$} & {E, MeV} & {} & 1702 & 1702 & 1711 & 1688
& 1702 & 1702 \\ {} & {$\Gamma$, MeV} & {} & 30 & 150 & 145 & 162 & 150 &
30 \\ \hline \hline \end{tabular} \label{tab:cluster34} \end{table}

For now, we did not calculate coupling constants in the 3-channel
approach, because here rather much variants of combinations of the resonance
cluster types are possible and not all the ones are considered to choose the
better variant, though already a satisfactory description is obtained. But
according to cluster types, we can conclude something about
the nature of the resonances considered. Our 3-channel conclusions
generally confirm the ones on the basis of the 2-channel analysis, besides
the surprising conclusion about the $f_0(980)$ nature. It turns out that
this state lies slightly above the $K\overline{K}$ threshold and is
described by a pole on sheet II and by a shifted pole on sheet III under
the $\eta\eta$ threshold without an accompaniment of the corresponding
poles on sheets VI and VII, as it was expected for standard clusters. This
corresponds to the description of the $\eta\eta$ bound state.

\noindent{\bf 3. Summary.}
On the basis of a combined description of the processes
$\pi\pi\to\pi\pi,K\overline{K}$ in the channel with $I^GJ^{PC}=0^+0^{++}$
with the parameterless representstion of the $\pi\pi$ background, a
model-independent confirmation of the $\sigma$-meson below 1 GeV is
obtained once more.

The existence of $f_0(600)$ with the properties of the $\sigma$-meson and
the obtained $\pi\pi$-scattering length ($a_0^0(\pi\pi)\approx 0.27$)
suggest the linear realization of chiral symmetry.

The 2-channel analysis evidences that $f_0(980)$ and especially
${f_0}(1370)$ (if exists) have the dominant $s{\bar s}$ component. From
the combined analysis of the processes
$\pi\pi\to\pi\pi,K\overline{K},\eta\eta$, we obtain an additional
indication for $f_0(980)$ to be the $\eta\eta$ bound state.

The best combined description of the processes
$\pi\pi\to\pi\pi,K\overline{K}$ is obtained with the states: $f_0(600)$,
$f_0(980)$, $f_0 (1500)$, and $f_0 (1710)$ (without ${f_0}(1370)$). The
$K\overline{K}$ scattering length is very sensitive to whether this latter
state exists or not.

Since $f_0(1500)$ has approximately equal coupling constants with the
$\pi\pi$ and $K\overline{K}$ systems, we conclude about its dominant
glueball component. This is confirmed by our 3-channel analysis where $f_0
(1500)$ is represented by the pole cluster corresponding to a flavour
singlet, {\it e.g.}, glueball.

In the 2- and 3-channel considerations, $f_0 (1710)$ is represented by the
pole cluster corresponding to a state with the dominant $s{\bar s}$
component.

\end{document}